# Quantitative strain analysis of InAs/GaAs quantum dot materials

Per Erik Vullum[1,2], Magnus Nord[2], Maryam Vatanparast[2]*, Sedsel Fretheim Thomassen[2], Chris Boothroyd[3], Randi Holmestad[2], Bjørn-Ove Fimland[4] and Turid Worren Reenaas[2]

[1] *Materials and Chemistry, SINTEF, NO-7465 Trondheim, Norway*
[2]*Department of Physics, Norwegian University of Science and Technology- NTNU, NO-7491 Trondheim, Norway*
[3]*Ernst Ruska-Centrum und Peter Grünberg Institut, Forschungszentrum Jülich, D-52425 Jülich, Germany*
[4]*Department of Electronics and Telecommunications, Norwegian University of Science and Technology- NTNU, NO-7491 Trondheim, Norway*

*maryam.vatanparast@ntnu.no

## Abstract

Geometric phase analysis has been applied to high resolution aberration corrected (scanning) transmission electron microscopy images of InAs/GaAs quantum dot (QD) materials. We show quantitatively how the lattice mismatch induced strain varies on the atomic scale and tetragonally distorts the lattice in a wide region that extends several nm into the GaAs spacer layer below and above the QDs. Finally, we show how V-shaped dislocations originating at the QD/GaAs interface efficiently remove most of the lattice mismatch induced tetragonal distortions in and around the QD.

## Introduction

Quantum dots (QDs) are nano-objects exhibiting 3D quantum confinement of charge carriers [1]. The atomic-like properties emerging from these systems have been exploited for advanced electronics and optoelectronics, such as light emitting devices and laser diodes [2,3], to intermediate band solar cells [4]. Self-assembled QDs are formed during the course of molecular beam epitaxy (MBE) on the GaAs surface by self-organization according to the Stranski-Krastanow (SK) mechanism and are then overgrown by a GaAs layer [5]. Beyond a critical thickness of the deposited film, QDs form spontaneously due to the lattice-mismatch strain between the QD and substrate materials. Structural aspects, such as the morphology and the elastic strain influence the optical properties of the dots. Therefore, understanding the physics that correlate strain, morphology, and composition of QDs is essential for the development of high quality and uniform QDs. A major challenge for the realization of intermediate band solar cells is the degradation of the QD and matrix structure due to the generation of misfit dislocations [6, 7]. Such defects are introduced as a result of residual compressive strain that accumulates as successive QD layers are grown [8, 9]. Hence, a detailed knowledge and control of the strain in QD-based devices are of major importance for their performance.

Precise experimental determination of the strain in buried QD systems is challenging. Most findings therefore describe the strain in a qualitative rather than quantitative manner due to the lack of resolution to quantify it on the atomic scale [10-12]. Numerical simulations can possibly provide quantitative strain data with high accuracy [13], but the results depend on a number of material system input parameters, parameters that usually are unknown or known with limited accuracy. However, with the development of aberration-corrected transmission electron microscopes the resolution has now reached the sub Å level, and this improved resolution opens the possibility for direct measurements of lattice strain on the atomic scale. These measurements are based on the assumption that there is a constant relationship between the intensity maxima in the (scanning) transmission electron microscopy ((S)TEM) images and the location of the atomic columns. This relationship gives a spatial shift of the intensity maxima positions with respect to the atomic columns [14]. Geometric phase analysis (GPA) [15] has proven to be a reliable method to determine the strain in atomic resolution images and from QD systems in particular [16-18]. In the present work, we use aberration corrected STEM in combination with GPA to quantitatively determine the strain in and around QDs.

Recently, a few groups have performed GPA on high resolution STEM images from QD systems. However, these findings have been from a system with large, facetted QDs [19], from materials with a different chemical composition than InAs/GaAs [20], or strain analyses on images with sub-optimal spatial resolution [21]. We here study materials that are more relevant for intermediate band solar cell applications, where the dots are typically small and lens shaped. In addition, we also show why dislocations originate from some QDs and we demonstrate how these dislocations significantly modify the strain fields in and around the dots.

## Materials and Method

The samples were grown in a Varian Gen II Modular MBE system with a dual-filament Ga source, a SUMO In source, and a valved cracker As source from Veeco. Si-doped GaAs (001) 2 quarter wafers were baked at 615°C for 10 minutes to remove the oxide. Two different samples were grown. In both samples, a GaAs buffer was grown at 586°C before the substrate temperature was lowered and stabilized at 500°C (hereafter called the "AlAs capped sample") or 510°C (hereafter called the "GaAs capped sample"). 2.77 monolayers (ML) (AlAs capped sample) or 2.35 ML (GaAs capped sample) InAs were deposited in cycles. During each cycle, the In shutter was open for 1 second and closed for 2 seconds. Both samples were continuously flushed with $As_2$. The InAs growth rate was 0.1 ML/s and the In/As flux ratio (i.e. beam equivalent pressure ratio) was 1:29. In the AlAs capped sample, the QDs were capped with 3 ML of AlAs under excess $As_2$ and at a substrate temperature of 500°C. The AlAs capped sample consists of three QD layers separated by 85 and 50 nm GaAs spacers. The GaAs capped sample consists of two QD layers separated by a 50 nm GaAs spacer. The first 10 nm of the spacers were grown at 500°C (AlAs capped sample) or 510°C (GaAs capped sample) and the final part of the spacers at 586°C. In both samples, the last QD layer was left uncapped on the surface, and the substrate temperature was ramped down towards room temperature immediately after the last QD layer was grown. The large spacer thicknesses were chosen to avoid any coupling between the QD layers.

Cross-section TEM samples were prepared by mechanical grinding, followed by dimpling and Ar-ion milling in a Gatan Precision Ion Polishing System (PIPS) using liquid nitrogen cooling. The acceleration voltage was initially 3.5 kV before gradually reduced to 1.5 kV in the final stages of the milling. The combination of liquid nitrogen cooling and a low acceleration voltage during milling is important in order to minimize sample damage. The TEM characterizations were performed with 3 different microscopes: low resolution images were acquired with a JEOL 2010F, operating at 200 kV. High resolution TEM images were acquired with a 300kV FEI Titan with a Cs imaging corrector. High resolution, high angle annular dark field (HAADF) STEM images were acquired with a 300kV FEI Titan with a Cs probe corrector. All TEM images were taken with the electron beam parallel to the crystallographic [110] direction.

## Results and discussion

Low magnification TEM images from both samples are shown in Fig. 1. The QDs are lens shaped in both samples, with a width typically in the range 22-26 nm and a height of 6-7 nm (AlAs capped sample) or 4-5 nm (GaAs capped sample). The increased QD height in the AlAs capped sample is due to a thicker InAs layer in the AlAs capped compared to the GaAs capped sample, but also due to the use of Al in the capping material. The Al atoms have a lower mobility than the Ga atoms and give less segregation of In into the capping or spacer material [22]. In addition, AlAs capping has been shown both to increase the height and the density of dots compared to GaAs capping [23]. No defects in the form of dislocations can be observed in the GaAs capped sample. In the AlAs capped sample, however, a few V-shaped dislocations can be observed. Such V-shaped dislocations are shown in Fig. 1 (a) and (c). These dislocations always initiate at the interface between a QD and the surrounding matrix and they terminate at the QD layer above. A few nm above the QD, the dislocations develop into a pair of pure 111 type of stacking faults, as seen in Fig. 1 (c). I.e. the dislocation on the left turns into a stacking fault and the one on the right side a stacking fault. A complete 3D pyramid of stacking faults exists if similar stacking faults are present along the two <111> directions that cannot be seen with the present crystal orientation [110] directions.

High resolution TEM images from QDs are shown in Fig. 2(a) (AlAs capped sample) and Fig. 2(c) (GaAs capped sample). The corresponding Fourier Transforms (FTs) are shown in Figs. 2(b) and 2(d). Perpendicular to the growth direction (parallel to the [1-10] directtion), the peaks in the FTs are perfect Gaussians, and no shoulders or extra intensity can be seen due to lattice parameter variations in this direction. Hence, the large lattice mismatch between the QDs and the surrounding GaAs matrix is elastically absorbed and the QD lattice parameter is compressed to match the lattice parameter of GaAs. However, along the growth direction (equal to the [001] direction) the strain profile is complex. In the FT of the AlAs capped QDs (Fig. 2(b)) significant intensity is present on the high $d$-value side of the major GaAs peaks. By looking closely at the blue intensity profile in Fig. 2(e), a small shoulder is also present on the low $d$- value side of the (004) peak. These two features indicate that a significant part of the crystal is exposed to a tensile strain, and a minor volume is compressively strained (relative to a cubic, unstrained GaAs lattice). This means that the crystal is tetragonally distorted with a c/a ratio > 1 inside the volume that gives the high $d$-value peak in the FT. A minor volume is tetragonally distorted with a c/a ratio < 1, corresponding to the region responsible for the low d-value shoulder of the (004) peak. In the FT of the GaAs capped QDs (Fig. 2(d)), high d-value peaks can also be observed parallel to the [001] direction. However, the intensity profiles along the [001] direction (see

Fig. 2(e)) show no detectable low *d*-value peak or shoulder, and the high d-value peak is less intense than in the FT from the AlAs capped sample.

The high resolution images and corresponding FTs shown in Fig. 2 are not able to correlate the tetragonal distortions observed in reciprocal space to the matching tetragonally distorted regions in real space. GPA on the other side is able to correlate strain features observed in the FT to specific atomic columns in real space. In order to avoid potential phase shifts of the atomic columns due to variations in sample thickness or due to changes in chemical composition, GPA has been performed on high-resolution STEM images. No noise or drift compensations were performed to any of the high resolution STEM images. Therefore, only strain parallel to the fast scanning direction was quantified in an image. Two images, one with the fast scanning direction parallel to the [110] direction, and a second image with the fast scanning direction parallel to the [001] growth direction, were acquired for every analyzed QD. The first image allowed quantification of the strain parallel to the [110] direction and the second image was used to quantify strain along the [001] direction. High resolution HAADF STEM images from QDs in the AlAs and GaAs capped samples are shown in Fig. 3. Two QDs are shown from the AlAs capped sample. These dots correspond to the dots labelled 2 (Fig. 3(a)) and 5 (Fig. 3(e)) in Fig. 1(c). Strain maps corresponding to each of the high resolution STEM images are also shown in Fig. 3. Only strain parallel to the [001] direction is shown in the maps. Maps that display the strain along other directions than [001] are not shown, since strain above the noise level cannot be detected in any of these maps. These strain results confirm the observations in the FTs in Fig. 2. An unmodified GaAs region away from any QD was used as a reference to define zero strain. Compressive and tensile strains are always described and compared to this unstrained GaAs reference.

The strain profile, [001], across QD 2 in Fig. 1(c) is shown in Fig. 3(g) and shows three distinct features: 1) A tensile strain that varies between 3 6 % is present inside a 6 nm wide region that corresponds to the region of the dot itself. 2) Just under the QD/GaAs baseline interface the GaAs lattice is exposed to a 3 % compressive strain. By moving away from the interface, this compressive strain decreases to zero over a distance of ca. 5 nm. 3) Above the QD the GaAs matrix is also exposed to a compressive strain. The compressive strain above the QD is smaller than the compressive strain below the QD, but the crystal above the QD seems to relax back to zero strain slower than in the region below the QD. In summary and within our resolution of 0.4 % strain, we here observe that the lattice is tetragonally distorted inside a ca. 20 nm wide region. Inside the QD, the lattice is tetragonally distorted with a c/a ratio of 1.03 1.06. Below and above the QD, the surrounding GaAs matrix is tetragonally distorted with a c/a ratio in the range 0.97 1.00. Relating these strain observations to intermediate band solar cell applications, a lower limit for the distance between each of the QD layers would be 20 nm with the present QD size and chemical compositions. Below this critical limit, strain will accumulate from one QD layer to the next. The critical limit is likely to be somewhat larger than 20 nm since we are not able to measure the strain with absolute precision.

The strain profile across the QD in the GaAs capped sample (see Figs. 3(d) and 3(g)) is significantly different from the strain profile in and around the AlAs capped dot. The dot height is just below 5 nm, compared to 6 nm for the AlAs capped dot. The tensile strain inside the dot varies from ca. 3 5 %, which is approximately in the same range as for the AlAs capped dot. The compressive strain in the surrounding GaAs however, is hardly visible with the present resolution. Some minor compression of the GaAs matrix can be observed in the first couple of nm above and below the dot. These results clearly show that the strain and accumulation of strain in multilayer QD structures do not scale linearly with the QD size. Moreover, it shows the importance of having a narrow size distribution of QDs without any large dots where strain can start to accumulate and possibly create threading dislocations.

The [001] strain map of the dot from which the V-shaped dislocations originate (Fig. 3(f)) shows some surprising results. An approximately 1 nm wide region running along the direction is tetragonally distorted. This distorted region is the remaining wetting layer (WL) which is inherently present in self-assembled StranskiKrastanow grown QD structures [24]. The TEM specimen is thicker than the diameter of the QD in Fig. 3(e) (which also can be seen as the AlAs capping is present as a sharp band crossing the entire cross-section of the QD) and a tetragonally distorted WL is therefore present across the dot. However, in the QD region [001] (relative strain in the [001] direction compared to the unstrained GaAs lattice) vanishes (Fig. 3(f)). QDs 4 and 6 in Fig. 1(c) are located in a region with the same TEM specimen thickness as QD 5 and these two QDs show strain map that are similar to the strain around QD 2. Hence, the "strain relaxation" seen in Fig. 3(f) is not an artificial effect of the TEM specimen thickness. The V-shaped dislocations are essentially two additional atomic planes, i.e. two edge dislocations, in the GaA matrix, symmetrically originating from each side of the QD. These two additional planes give a net expansion of the GaAs matrix compared to the QD, which does not contain these additional planes, by 1.88 and 5.33 along the [001] and directions, respectively. Hence, a significant portion of the strain (relative to the unstrained GaAs lattice) is compensated for by the two additional planes originating at the QD/GaAs interface. Integrating the strain profile of QD 2 over the 6 nm height of the dot (green line in Fig. 3(g)) gives a total tensile strain of 2.2. These simple calculations clearly show that most of the strain relative to unstrained GaAs is expected to relax by the insertion of an extra atomic layer at each side of the QD/GaAs interface. As such, the strain profile across QD 5 is not surprising.

The present TEM characterization based on aberration corrected TEM and STEM images of embedded InAs/GaAs QDs shows quantitatively how the strain varies in two dimensions and with atomic column resolution. We here show that all strain is elastically absorbed perpendicular to the growth direction, i.e. in the (001) plane. Parallel to the growth direction, i.e. parallel to the [001] direction, the QDs are tetragonally stained with a c/a-ratio ¿ 1 in the dots and with a c/a-ratio ¡ 1 in the GaAs above and below the dots. Finally, we show how V-shaped dislocations originating at the QD/GaAs interface efficiently reduce most of the elastic strain relative to the unstrained GaAs lattice.

## Acknowledgements

This project was funded by the research council of Norway under contract no. 181886. Centre for Electron Nanoscopy, Denmark Technical University, is acknowledged for providing access to probe and image Cs corrected Titans.

## Author Contributions

Per Erik Vullum, Randi Holmestad, Bjørn-Ove Fimland and Turid Worren Reenaas designed the study and planned the experiments, Sedsel Fretheim Thomassen grew the samples, Magnus Nord made TEM samples, Per Erik Vullum, Magnus Nord and Chris Boothroyd did the experiments and collected the data, Per Erik Vullum, Magnus Nord and Maryam Vatanparast performed the analysis, and all authors contributed in scientific discussions and in writing the manuscript.

## Competing financial interests

All authors declare no competing financial interests.

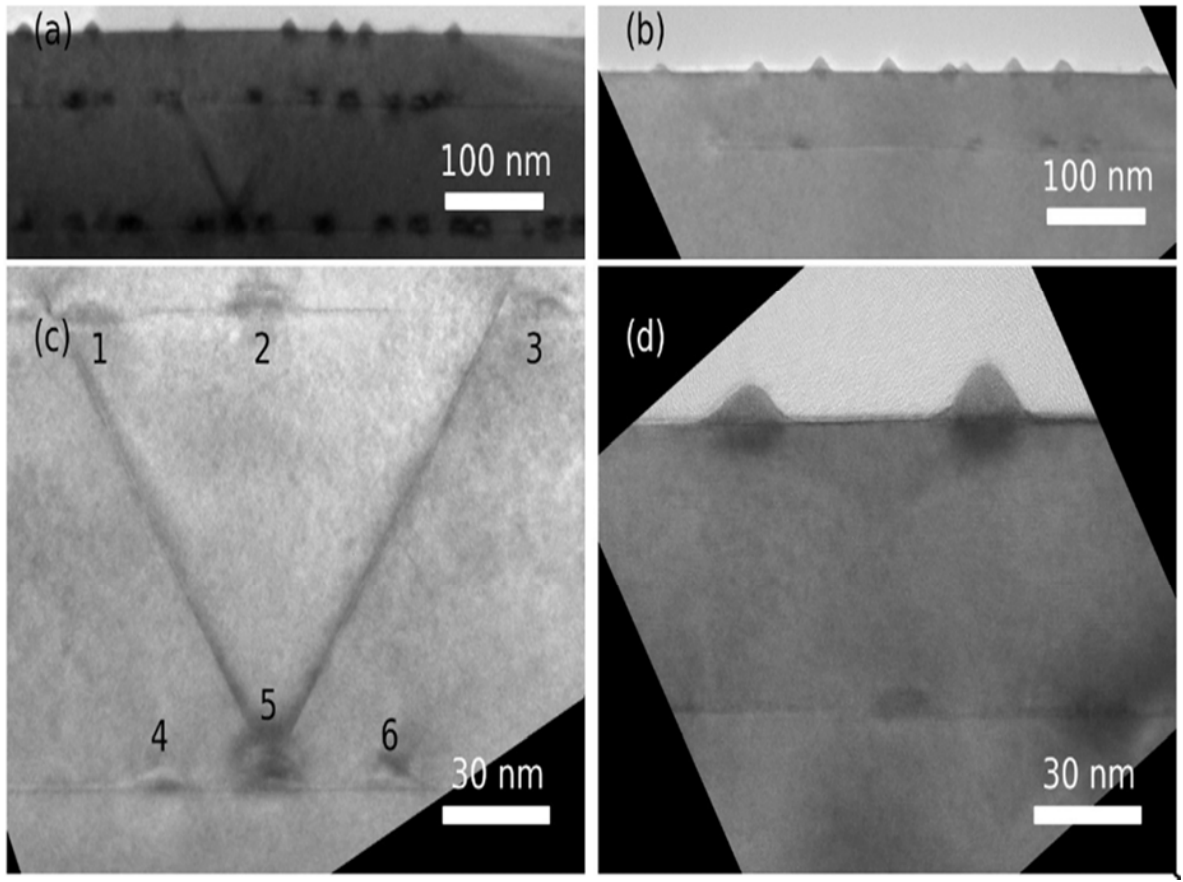

**FIG. 1.** Low resolution BF TEM images from the AlAs capped (a + c) and GaAs capped (b+ d) samples. Isolated QDs are shown with medium magnification in Figs. (c) and (d). In Fig. (c) a V-shaped dislocation originates from a QD in the first layer, and each of the dislocation lines terminate at a QD in the second layer.

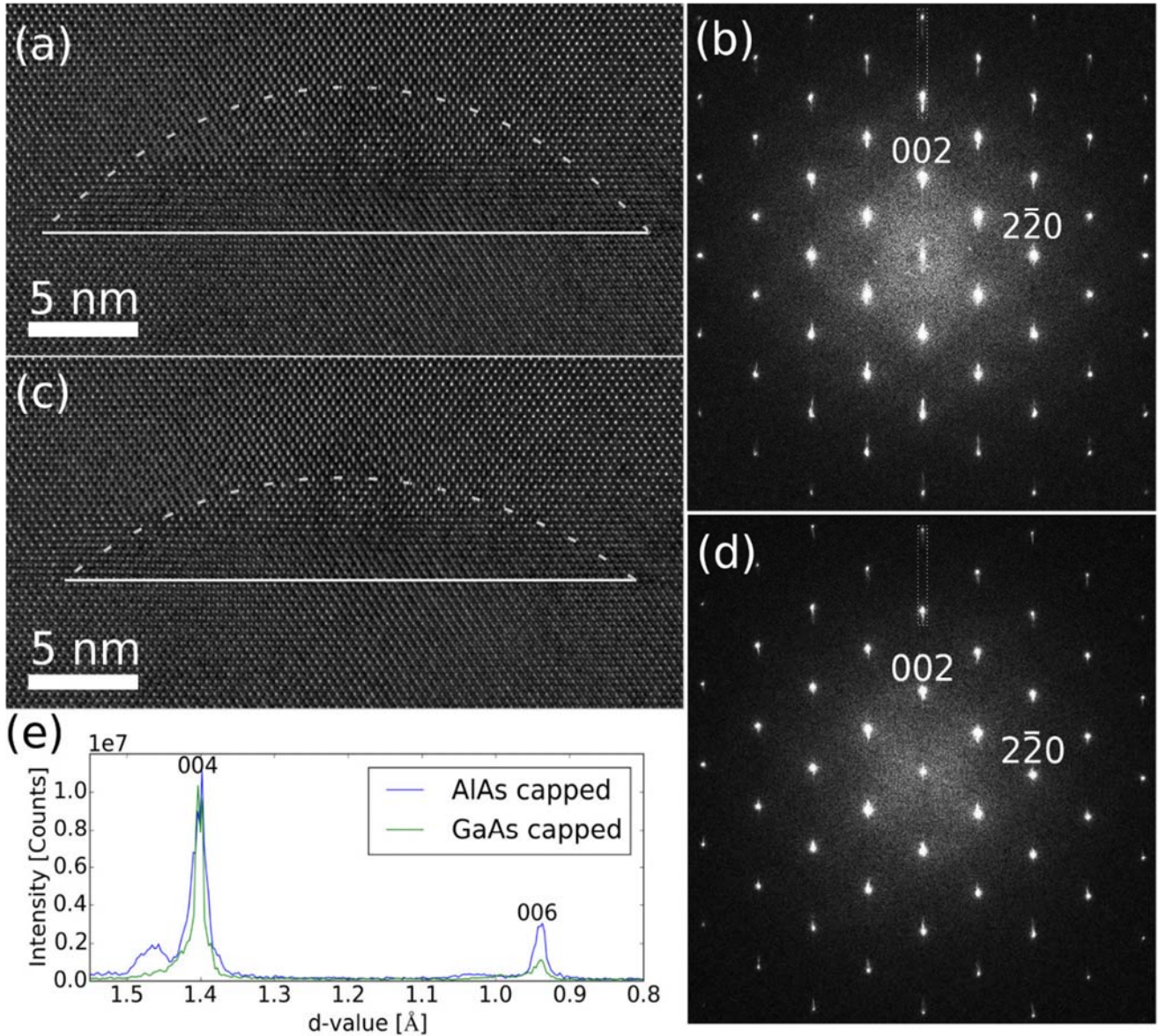

**FIG. 2**. High resolution TEM images of embedded QDs in (a) the AlAs capped and (c) the GaAs capped samples. The corresponding Fourier transforms (FTs) are shown in (b) and (d). The graph in (e) shows the intensity in the FTs parallel to the [001] growth direction, covering the (004) and (006) peaks. This area is marked with dashed lines in each of the two FTs.

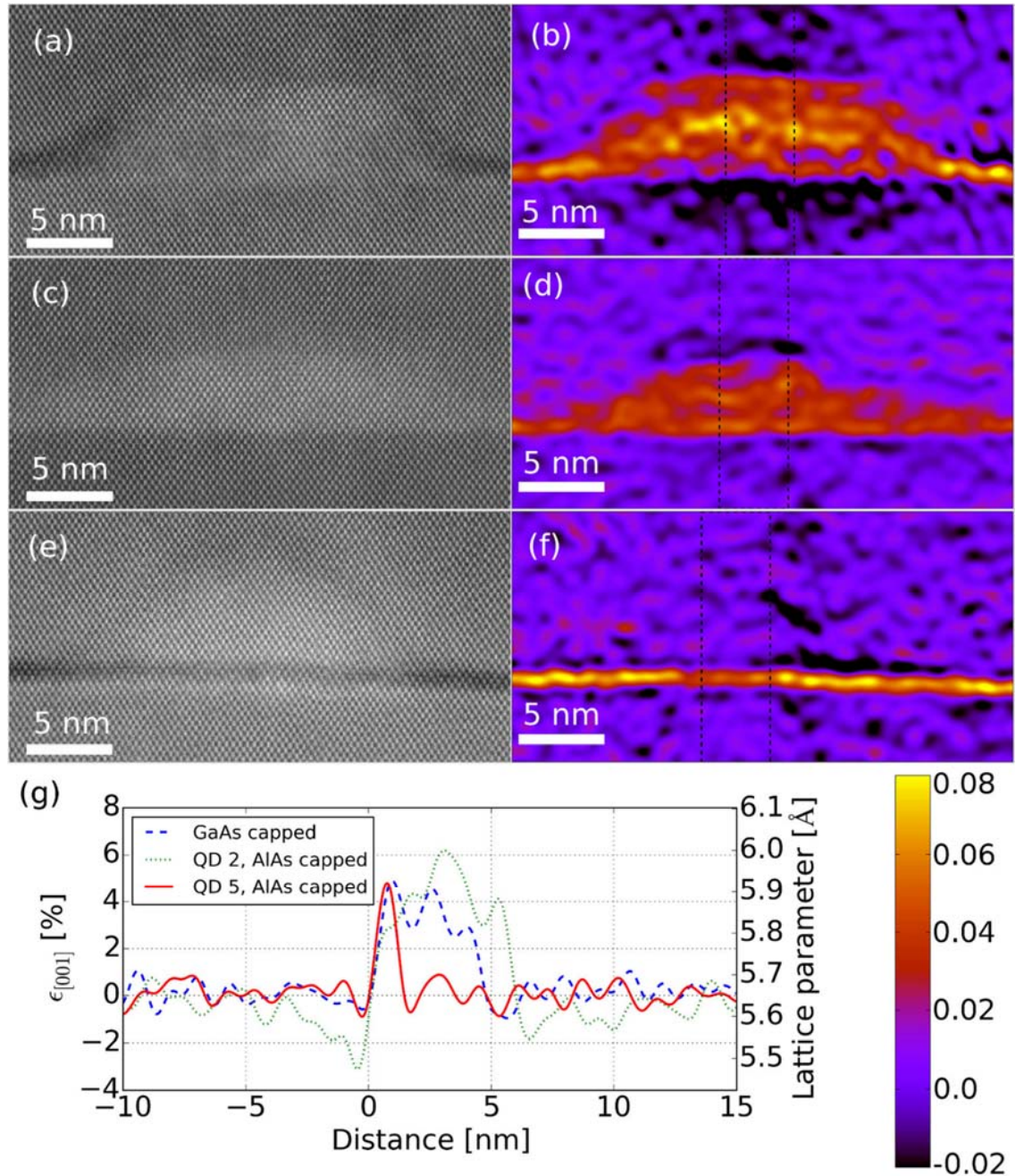

**FIG. 3.** High resolution HAADF STEM images of QDs from the AlAs capped sample, (a) and (e), and the GaAs capped sample, (c). The QDs shown in (a) and (e) are the two dots labelled 2 and 5, respectively, in Fig. 1 (c). Figs (b), (d) and (f) show strain maps for strain parallel to the crystallographic [001] direction. Quantitative values for the strain along a line crossing through the middle of the QDs are given in Fig. 3(g). Zero strain is defined in an unstrained GaAs region away from any QD, and distance "0" set at the interface between the QD and the GaAs matrix below.